
\overfullrule=0pt
\magnification=\magstep1
\baselineskip=3ex
\raggedbottom
\def\sk {\vskip 10pt}

\catcode`@=11
\def\om {\omega}
\def\pa {\parallel}
\def\ra {\rightarrow}
\def\la {\lambda}

\def\ep {\epsilon}
\centerline{\bf Non-Fermi Liquid Behavior in $3+1$ Dimensions with PT
Invariant}
\vskip 5pt
\centerline{\bf Gauge Field: An Renormalization Group Approach}
\bigskip
\centerline{ Gu Yan}
\sk
\centerline{\it Department of Physics, Fudan University}
\centerline{\it Shanghai 200433, PRChina \footnote{$^{*}$}{current address:
Department of Physics, University of Pennsylvania, Philadelphia, PA 19104,
U.S.A.}}
\bigskip
\bigskip
\bigskip
\centerline{\bf Abstract}
\vskip 1.5cm
We introduce a Hamiltonian coupled between a normal Fermi surface and a
polarized Maxwell type gauge field.We adopt a {\it calibrated scaling }
approach in order to be consistent with the results obtained at $2+1$
dimensions as well as the Weinberg's theo
rem.
Renormalization group equations are calculated by using a dimension regulator.
It turns out that the gauge interaction is controlled by a non-trivial fixed
point.
The two-point Green's function can thus be calculated accurately in the low
temperature limit and possesses a non-quasi particle pole which signals the
break down of the Landau Fermi liquid theory.
In addition,we show that the anomalous dimension of fermion field is in general
gauge dependent. But it will survive under generic gauge choice due to the
renormalization of the gauge fixing parameter.
\bigskip
\medskip
\sk
It is well believed that both the normal state of high $T_c$ superconductor and
fractional quantum Hall effect demonstrate non-Fermi liquid behavior which
cannot be understood by the usual Landau Fermi liquid theory [1][2]. A natural
candidate to realize
non-Fermi liquid behavior is to include gauge interaction. Besides the
advantage that gauge fields enhance low energy interactions while remain
insensitive to a fine tune of the shape of Fermi surface [3], it is more
accessible to perturbative theories in
 which renormalization group (RG) approach appeals to be a powerful tool.
In $2+1$D, if PT invariance is spontaneously broken, Chern-Simon field seems to
be an attractive choice, and interesting and fruitful results have been
obtained recently by using a RG approach [4].
On the other hand, if PT invariance is preserved, the Maxwell action should be
the right choice, which is also favored experimentally [6].
Non-Fermi liquid behavior with PT invariant gauge field at $2+1$D have been
broadly explored[3][7][8].
But since the relevant coupling might not be small as we shall see, those
results obtained through a usual perturbation scheme might be inconclusive at a
zero temperature limit.
\sk
In this paper, we extend our focus to $3+1$D and work with a relatively simpler
model where the divergences are more tractable and therefore may presumably
cast some light on the common features of the PT invariant gauge fields.

\sk
One way to study the non-Fermi liquid behavior is by calculating the 2-point
Green's function $G(\om, k)=<\psi(\om,k)^{\dagger}\psi(\om,k)>$. For a normal
Landau Fermi state, $G(\om,k)$ has the form ${1\over
i\om-\epsilon(k)-\Sigma(\om,k)}$ where the
self energy $\Sigma(\om, k)$ must behave as $\Sigma\sim (\om-\mu)^2$ to assure
quasi particle poles.
However, if the retarded Green's function has a branch cut instead of a single
pole the system will develop non-Fermi liquid behavior, e.g, the singularity at
Fermi surface and anomalous low temperature behavior which cannot be accounted
for by the Fermi
liquid theory.
\sk
In the following sections, we first formulate the problem by introducing an
effective Maxwell type gauge interaction to the usual Fermi liquid state in
$3+1$D. Next, by assuming Weinberg's theorem as well as to be consistent with
the results obtained in $
2+1$D, we explain the {\it calibrated scaling} approach adopted in this paper.
And then, we use a dimension regulator to calculate the counterterms  of the
renormalized Hamiltonian as well as the RG equations.
In particular, we find the gauge dependence of the anomalous dimension of the
fermion field and discuss the consequences.
Finally, we rescale the Green's function by the anomalous dimension and explain
the resultant branch cut singularities in the two-point Green's function which
indicates the non-Fermi liquid behavior of the system.
We briefly mention the generalizations at the end.
\sk
To begin with, we consider the following Hamiltonian in $3+1$D with proper
renormalizations.
$$
S=\int d\om d^3 k Z_{\psi}\psi^{\dagger}(\om,k)(i\om-\ep(k))\psi(\om,k)+\int
d\om d^3 kZ_{A_0}A_0(\om,k)(-\om^2)A_0(\om,k)
$$
$$
+2\sum_i \int d\om d^3 k Z_{A_0}^{1\over 2}Z_{A_i}^{1\over 2}A_0(\om,k)\om
k_i(1-\la)A_i(\om,k)
$$
$$
+\sum_{i,j}\int d\om d^3k Z_{A_i}^{1\over 2}Z_{A_j}^{1\over
2}A_i(\om,k)[k^2+k_ik_j(1-\la)]A_j(\om,k)
$$
$$
+ig\mu^{\ep}\int d\om d^3 k d\om'd^3k'Z_gZ_{\psi}Z_{A_0}^{1\over
2}\psi^{\dagger}(\om+\om',k+k')\psi(\om,k)A_0(\om',k')
$$
$$
-g\mu^{\ep}\int d\om d^3k d\om'd^3k'Z_gZ_{\psi}Z_{A_i}^{1\over
2}\psi^{\dagger}(\om+\om',k+k')\psi(\om,k)A_i(\om',k'){\partial\over \partial
k_i}\ep(\om+\om',k+2k')
$$
Where $\ep(k)=k_y+\beta(k^2_x+k^2_z)$, $\beta={1\over K_F}$. $Z_{\psi}$,
$Z_{A_i}$ and $Z_{g}$ are the renormalization coefficients of the fermion field
${\psi}$, gauge field $A_i$ and gauge coupling constant $g$. We have taken
$v_F=1$ for simplicity. The
 reason will be explained later.
The first term is the free spinless fermion (will be referred to as fermion
since now) field, the second and the third terms are the free gauge field with
gauge fixing parameter $\la$. The last two terms describe the gauge interaction
which dominates in t
he direction parallel to the Fermi surface as a requirement of gauge
invariance.
\sk
We need to determine the scaling behavior of the momenta, fields and gauge
coupling constant before we can compute the loop integrals correctly.
\sk
We adopt a calibrated scaling approach which is based on the following
considerations:
\sk
(1). Since we consider a RG transformation that scales the Hamiltonian toward a
single point on the Fermi surface, the scaling of $k_y$ and $\om$ might not
necessarily be the same as the scaling of $k_x$ and $k_z$ [4].
 If $k_y\rightarrow Sk_y$, $\om\ra S\om$, we can in general write the scaling
of $k_x$ and $k_z$ as $k_x\ra S^{\gamma}k_x$ and $k_z\ra S^{\gamma}k_z$.
Consequently, the scaling of the gauge coupling g should be $g\ra S^{1-{1\over
2}(d-2)\gamma}g$, where d
 is the number of dimension.

(2). It has been proved by using a bosonization technique that Coulomb
interaction at $2+1$D will leave the Fermi liquid picture unchanged [9]. Since
the unscreened Coulomb interaction can be considered as the tree approximation
of the above gauge interac
tion, we can infer that the Coulomb interaction vertex must be either marginal
or irrelevant at tree level.

(3). In ref. [10], it is shown that at $2d+1$D where $1<D\leq2$, for long range
4-point interaction $r^{-(2-d)}$, where d can be arbitrarily close to but not
equal to 1 which corresponds to the Coulomb interaction, the Fermi liquid
picture breaks down.
\sk
Based on the above considerations, we can conclude that the Coulomb interaction
vertex must be marginal at tree level in $2+1$D. And therefore g must be scaled
as $g\ra S^{1\over 2}g$ at tree level. The corresponding value for $\gamma$ is
1, meaning that
for PT invariant gauge field, $k_x$, $k_y$, $k_z$ and $\om$ are be scaled
identically.
Hence we should have $\beta\ra S^{-1}\beta$, meaning the quadratic term in
kinetic energy is irrelevant at tree level. This result is also in agreement
with the Weinberg's theorem as we'll see later.
However, as we move our attention to one loop level later, we'll find that
$\beta$ is a dangerous irrelevant parameter (similar term is coined in ref. [4]
but under a different context) which can not be simply taken as zero at the
beginning. The other way
 to understand this is, suppose $\beta$ is set zero, then the gauge interaction
and consequently the Hamiltonian as a whole can be effectively reduced to a
$1-$D model which is obviously incorrect.
As a matter of fact, $\beta(k^2_x+k^2_z)$ is the leading term which causes the
gauge interaction between two fermions around the same point near the Fermi
surface.
Before it is scaled to zero, it might cause some relevant effect (e.g.
divergences) at higher loop order.
Thus it can only be taken as zero when it is subleading in the calculation.
\sk
Now, we're ready to calculate the loop integrals. At one loop order, we
consider 3 graphs (fig.1), namely, the the fermion self-energy graph, the
vacuum polarization graph and the vertex correction graph. We need only to
calculate two of these graphs by t
he virtue of the following Ward identity
$$
\eqalign{
& -\la(\partial^2_t-\partial_i\partial_i)<T\partial\cdot
A(z)\psi(y)\psi^{\dagger}(x)>\cr
&=g<T\psi(y)\psi^{\dagger}(x)>\delta(z-y)-g<T\psi(y)\psi^{\dagger}(x)>\delta(z-x)\cr
}
$$
which can be easily proved by assuming the general Green's function identity.
If we cancel out the divergences on both sides, we'll get
$$
Z_A^{1\over 2}Z_{\psi}Z_A^{-1}=Z_gZ_{\psi}
$$
where $Z_A^{-1}$ comes from the absence of a counterterm for gauge mixing. So
that$$
Z_g=Z_A^{-{1\over 2}}
$$
which should be true for all gauge choices.
\sk
We apply a dimension regulator to handle the divergences which has the
advantage of preserving gauge invariance and being effective for both infra and
UV divergences. The calculation can be illustrated by carrying out the loop
integrals in fig. 2, the vac
uum polarization graph as shown below.
$$
\Gamma={g^2\over (2\pi)^4}\int d\om d^3k{1\over i\om-\ep(k)}{1\over
i(\Omega+\om)-\ep(k+K)}
$$
where k and K are vectors. First of all, as argued in [3], $K_y$ is much larger
than the other momentum components. As a result, although the excitations might
be close to the Fermi surface, the kinetic energy term
is actually very large so that $\Gamma$ does not yield a constant fermion
density as in ordinary Fermi liquid theory.
The other way to understand this is that we want the gauge symmetry to be
preserved which renders the gauge boson to be massless.
Next, we note that $\Gamma$ is invariant under the transformation $\delta\Omega
\ra\delta\Omega+i\delta E$, $\delta K_y\ra\delta K_y-\delta E$, so that
the counterterm proportional to $\Omega$ must be the same as the counterterm
proportional to $K_y$. To simplify the calculation, we can set $K_y=0$ and the
functions odd in $k_y$ disappear. Secondly,
during the the calculation, we'll just seek the leading terms in x and z
directions which absence might cause unphysical divergences as we shall see
below.
\sk
To begin with, we choose the origin to be on the Fermi surface so that
$k_y\geq0$. Then, we make a wick rotation and assume a positive metric. When
the counterterms are sorted out, we will analytically continue it back.
Hence,
$$
\Gamma={g^2\over (2\pi)^d}\int d\om d^3k {1\over \om+k_y+\beta
k_{\pa}^2}{1\over \om+\Omega+k_y+\beta(k_{\pa}+K_{\pa})^2}
$$
where the minus sign from the fermion loop is canceled by a minus sign from the
two gauge interaction vertices. Using the Schwinger representation of the free
operator [11], we get
$$
\eqalign{
 \Gamma & ={g^2\over (2\pi)^d}\int_{0}^{\infty}da\int_{0}^{\infty}db\int d\om
\d^3k
exp(-a\om-ak_y-a\beta k_{\pa}^2-b\om
 -b\Omega-bk_y-b\beta k_{\pa}^2 \cr
& -2\beta k_{\pa}\cdot K_{\pa}-b\beta K_{\pa}^2)\cr
& ={g^2\over (2\pi)^d}\int_{0}^{\infty} da \int_{0}^{\infty} db\int d\om d^3k
exp(-(a+b)\om-(a+b)k_y-\beta^2(a+b)k_{\pa}^2 \cr
& -b(\Omega+\beta K_{\pa}^2+2bk_{\pa}\cdot K_{\pa}) \cr
}
$$
Make  the transformation $k_{\pa}-{k_{\pa}b\over a+b}$ and change the variables
$z=a+b$, $x={b\over c}$,
$$
\eqalign{
\Gamma & ={g^2\over (2\pi)^d}\int_{0}^{\infty} da \int_{0}^{\infty} db\int d\om
d^3kexp[-(a+b)-(a+b)k_y-(a+b)\beta k_{\pa}^2-b\Omega] \cr
&={g^2 \over (2\pi)^d}\int_{0}^{1}dx\int_{0}^{\infty}dz z\int d\om d^3k
exp[-z(\om+k_y+\beta k_{\pa}^2)-zx\Omega] \cr
}
$$
As at tree level, we scale $\om$, $k_y$ and $k_{\pa}$ by z and $\beta$ by
$Z^{-1}$, then we get
$$
\Gamma={g^2\over (2\pi)^d}\int_{0}^{1} dx \int_{0}^{\infty}
dzz^{1-d}exp(-zx\Omega)\int d\om d^3k exp[-(\om+k_y+\beta k_{\pa}^2)]
$$
Now it's clear in order to make the last integral finite, we can't set
$\beta=0$ at the beginning, because it's the leading term in x and z
directions. Besides, the wick rotation we made before should actually be a sign
dependent one which is in the
same spirit as doing the contour integral. In addition, we note that the
scaling we adopted at tree level is also in agreement with the Weinberg's
theorem, namely, the counterterm of a $1PI$ graph must be polynomial in the
external momenta of the
superficial degree of divergence. In other words, the relative scaling of
$\beta$ and $k_{\pa}$ should obey this theorem in which case we find we do.
\sk
After carrying out the internal loop integral, we eventually get
$$
\Gamma={g^2\over (2\pi)^{d-1}\beta}\Gamma (2-d)\int_{0}^{1} dx(-x\Omega)^{d-2}
$$
which has a pole at:${g'^2\Omega^2\over 48\pi^3 \ep}$, where
$g'=g\beta^{-{1\over 2}}$ and $\ep=4-d$
Thus $Z_A=1-{g'^2\over 48\pi^3\ep}$.
\sk
The calculation of fig. 1 can be done similarly, and we get
$Z_{\psi}=1-(3+{1\over \la}){g'^2\over 24\pi^{5\over 2}\ep}$
By the virtue of the Ward identity, we get in addition $Z_g=Z_A^{-{1\over
2}}=1+{g'^2\over 96 \pi^3 \ep}$ and $Z_{\la}=Z_A^{-1}=1+{g'^2\over 48\pi^3
\ep}$
\sk
The corresponding renormalization group equations can be calculated by assuming
the bare Green's function is invariant under the RG transformation.
For instance, from $g'_0=Z_g g'\mu^{\ep+{1\over 2}(1-\ep)}$, we can derive
$$
\eqalign{
\mu{d\over d\mu}g_{0}'=0= & ({1+\ep \over 2})\mu^{1+\ep\over 2}(g'+{g'^3 \over
96\pi^3\ep}) \cr
& +\mu^{1+\ep \over 2}(\beta (g')+{3g'^2\over 96\pi^3\ep}\beta (g')) \cr
}
$$
So that at $d=4$,
$$
\beta(g')={dg'\over dt}= {1\over 2}g'-A_1 g'^3
\eqno (1)
$$
where $A_1={1\over 48\pi^3}$. By the same token, we get the other two RG
equations, namely
$$
\beta_{\la}(g')={d\la \over dt}=A_{1}\la g'^2
\eqno (2)
$$
$$
\gamma_{\psi}={d\ln Z_{\psi}\over dt}=-(3+{1\over \la})A_2g'^2
\eqno (3)
$$
where $A_2={1\over 12\pi^{5\over 2}}$.
\sk
The fixed point for $\beta$ function occur at
$$
g'^{*}={1\over \sqrt{2A_1}}
$$
 From (2), it's obvious that at the fixed point, as long as $g'(0)>0$, $\la \ra
\infty$ when $t\ra \infty$, which corresponds to a Landau gauge.
Meaning that the gauge choice at tree level is irrelevant because it will be
forced to run into the Landau gauge eventually.
The corresponding anomalous dimension of the fermion field is:
$$
\gamma^{*}=\gamma^{*}_{\psi}={3\over 2}{A_2\over A_1}=6\pi^{1\over 2}>0
$$
which is gauge independent after the renormalization of gauge fixing is taken
into account.
\sk
With the $\beta$ and $\gamma$ functions by hand, we can now calculate the two
point Green's function where most of the important physical properties (e.g.
response functions) come from. The renormalized and unrenormalized Green's
functions are related by
$Z_{\psi}$, the anomalous dimension of the field.
$$
\eqalign{
G_0(\om,k_y,k_{\pa},g_0,\epsilon) & =Z_{\psi}G(\om,k,k_{\pa},g(\mu),\mu)\cr
 & S^{-1}G(S\om,Sk_y,S^{1\over
2}k_{\pa},g(S\mu),\mu)exp[\int_{g(\mu)}^{g(S\mu)}
dg \gamma (g)/\beta (g)] \cr
& \sim {S^{\gamma^*}\over S (iw-\ep (k))} \cr
}
$$
The fermion Green's function in the zero temperature limit thus takes the form
$$
<\psi (k,\om) \psi^{\dagger}(k,\om)>\sim {({\om \over \mu_0})^{\gamma^{*}}
\over i\om-\ep(k)}
$$
Therefore, the quasiparticle pole $(i\om-\ep(k))G$ disappears since
$\gamma^{*}>0$.
Consequently, the massless particle which gives rise to the singularity cannot
be treated as a quasiparticle. Namely the Fermi  liquid picture breaks down due
to the long range gauge interaction.
\sk
Some of the above results can be generalized as follows:
\sk
(1). For a theory governed by a infra fixed point, the non-vanishing  anomalous
dimension of the Fermion field always leads to some kind of non-Fermi liquid
behavior. And the strength of the anomaly is characterized by the ratio between
the coefficients
of $\gamma$ and $\beta$ function, namely ${A_2\over A_1}$ (positivity of the
metric constrains $A_1$ to be positive [12]).
In general, $A_1$ is gauge dependent and we have shown for the gauge field
adopted in this  paper the gauge choice is irrelevant due to the
renormalization on the gauge fixing parameter.

(2). We did not actually need to assume a particular geometry of the Fermi
surface. This is because the contribution from the gauge interaction is
important only when
 two fermions are very close to each other in k space where the Fermi surface
is locally identical to a round one. Thus the above argument should hold
for Fermi surface with generic shape.
\sk
The author wishes to thank C. Nayak and F. Wilczek for ref. [5] before its
publication , and C. Kane for some useful discussions. \sk
\noindent
{\bf References}
\sk
\item{[1].}P.W. Anderson, Phys. Rev. Lett. {\bf 64}, 1839(1990); {\bf 65},
2306(1990); {\bf 66}, 3226(1991); {\bf 67}, 2092(1992)

\item{[2].}B.I. Halperin, P.A. Lee and N. Read, Phys.Rev. {\bf B 47},
7312(1993)

\item{[3].}J. Polchinski, "Low Energy Dynamics of the Spinon-Gauge System" ITP
preprint, NSF-ITP-93-93

\item{[4].}C. Nayak and F. Wilczek, Nuclear Phys. {\bf B 417}, 359(1994)

\item{[5].}C. Nayak and F. Wilczek, "Renormalization Group Approach to Low
Temperature Properties of a Non-Fermi Liquid Metal" PUPT 1489/IASSNS-HEP 94/59
\item{[6].}S. Spielman, J.S. Dodge, L.W. Lombardo, C.B. Eom, M.M. Fejer, T.
Geballe and A. Kapitulnik, Phys.Rev.Lett. {\bf 68}, 3472(1992)

\item{[7].}P.A. Lee, Phys. Rev. Lett. {\bf 63}, 680(1989)

\item{[8].}N. Nogaosa and P.A. Lee, Phys. Rev. {\bf B 43}, 1233(1991)

\item{[9].}A. Houghton, H.-J. Kwon, J.B. Marston and R. Shankar, J. Phys. {\bf
CM 6}, 4909(1994)

\item{[10].}P. Bares and X.G. Wen, Phys. Rev. {\bf B 48}, 8636(1993)

\item{[11].}J.C. Collins, {\it Renormalization}, Cambridge University Press,
1984.

\item{[12].}J.D. Bjorken and S.D. Drell, {\it Relativistic Quantum Fields},
McGraw-Hill, New york, 1966

\bye